\documentclass[aps,pra,twocolumn,groupedaddress,floatfix,showpacs]{revtex4-1}

\usepackage{graphicx}
\usepackage{dcolumn}
\usepackage{bm}
\usepackage{amsmath}
\usepackage{amssymb}
\usepackage{subfigure}
\usepackage{float}
\usepackage{multirow}
\usepackage{dcolumn}

\begin{document}

\title{Thermal noise of whispering gallery resonators}

\author{Akobuije Chijioke}
\author{Qun-Feng Chen}
\author{Alexander Yu. Nevsky}
\author{Stephan Schiller}
\affiliation{Institut f$\ddot{u}$r Experimentalphysik, Heinrich-Heine-Universit$\ddot{a}$t D$\ddot{u}$sseldorf, 40225 D$\ddot{u}$sseldorf, Germany}

\email{step.schiller@uni-duesseldorf.de} 


\date{\today}

\begin{abstract}
By direct application of the fluctuation-dissipation theorem, we numerically calculate the fundamental dimensional fluctuations of crystalline CaF$_2$ whispering gallery resonators in the case of structural damping, and the limit that this noise imposes on the frequency stability of such resonators at both room and cryogenic temperatures. We analyze elasto-optic noise - the effect of Brownian dimensional fluctuation on frequency via the strain-dependence of the refractive index - a noise term that has so far not been considered for whispering-gallery resonators. We find that dimensional fluctuation sets a lower limit of $10^{-16}$ to the Allan deviation for a 10-millimeter-radius  sphere at 5 K, predominantly via induced fluctuation of the refractive index.
\end{abstract}


\maketitle


\section{Introduction}

The properties of a body in equilibrium at a finite temperature undergo continuous fluctuation about their mean values. The fluctuation-dissipation theorem of Callen and Welton \cite{CallenWelton} established the relationship between these fluctuations and the dissipative properties of the system. This fundamental relationship is of great practical importance in precision measurements, such as the measurement and stabilization of the frequency of laser light using high-quality-factor optical cavities. Thermal noise sets a limit on the frequency stability of optical cavities, and hence on their performance in such applications.

The thermal noise of conventional Gaussian-mode optical cavities has been analyzed in detail \cite{Numata04, Levin, BGV99, BHV98}, providing guidance in the design of such cavities. A relatively-new type of high-finesse optical cavity is the whispering-gallery resonator (WGR) \cite{Chiasera}, in which light is confined inside an axially-symmetric dielectric object by continuous-total-internal reflection. Solid dielectric optical WGRs were quickly recognized as potentially useful \cite{Braginsky89} due to their high quality factors and small mode volumes, resulting in an intense confined coherent optical field. The lowest cavity loss demonstrated so far resulted in a linewidth of about 1 kHz in pure crystalline CaF$_2$ \cite{TenMillionFinesse}, making these resonators very attractive candidates for precision laser frequency stabilization. Understanding thermal noise in such resonators is critical for this application. For a WGR, both the fluctuation of the radius and of the refractive index must be considered. We present a calculation, by direct application of the fluctuation-dissipation theorem, of the frequency instability due to fundamental thermal fluctuation of the dimensions of crystalline CaF$_2$ resonators. 

Initial estimates of thermal noise in WGRs have been made \cite{MatskoEtAl_FundLimits, GorodetskyGrudinin}, in particular for thermorefractive noise, the fluctuation of refractive index due to fundamental temperature fluctuations. The thermorefractive noise has recently been measured for crystalline resonators at room temperature \cite{MPQ_MgF2_TRNoise}. The thermal noise due to fundamental dimensional fluctuations at a fixed temperature has been less investigated. For laser frequency stabilization, the fluctuations on timescales of 0.01 s to hundreds of seconds are of interest, for applications such as clock lasers for optical clocks \cite{NISTClockLaser} or tests of fundamental symmetries such as Lorentz invariance \cite{EiseleEtAl}. We find that at low temperatures the thermal noise due to dimensional fluctuation dominates over the thermal noise due to fundamental temperature fluctuation. We find that the dominant component of the dimensional-fluctuation thermal noise is due to induced changes of the refractive index, via the elasto-optical effect. 

To calculate the thermal noise we follow the direct approach introduced by Levin \cite{Levin} for standard two-mirror resonators, adapting it to WGRs. We perform the calculations numerically, by finite-element analysis, for sphere- and disk-shaped resonators. 


\section{Calculation}

\begin{figure}[!t] 
\begin{center}
\subfigure{\includegraphics[width=3.3in]{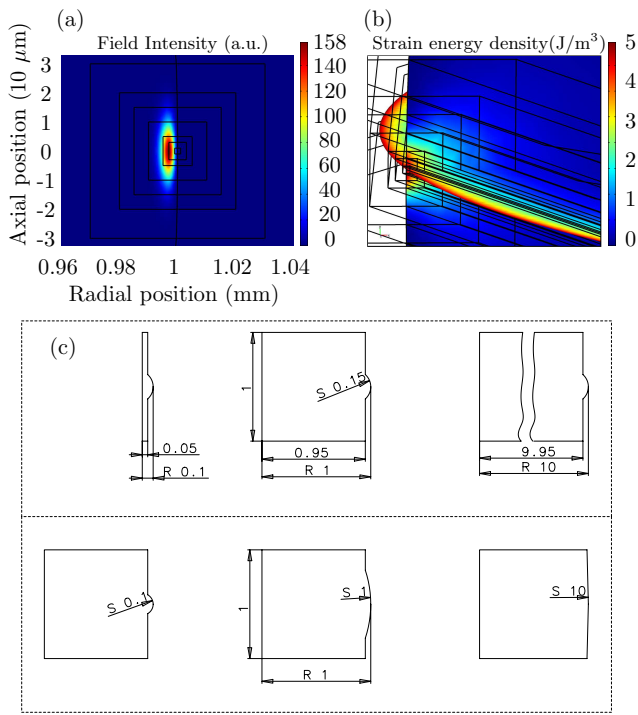}}
\caption{(a) Simulated mode profile of a  CaF$_2$ disk, as shown by the electric field intensity (arbitrary units). The radius ($R$), surface radius of curvature ($S$) and thickness are all 1 mm. (b) Simulated strain energy density distribution in the same disk. (c) Geometries used for disk simulation.}
\end{center}
\label{fig:Simulation}
\end{figure}

\subsection{Temperature fluctuation}

Our calculations reported in this paper are focused on the fundamental dimensional fluctuations that occur at a fixed temperature. However, to determine the expected dominant thermal noise terms in different regimes and to make contact with existing thermal noise results for WGRs, we compare our calculated noise levels to thermal noise arising from fundamental temperature fluctuations at a given mean temperature. These cause WGM frequency fluctuation via the dependence of refractive index on temperature (thermorefractive noise, TR) and thermal expansion of the resonator (thermoelastic noise). We evaluate the thermorefractive noise of a sphere using the expression of Gorodetsky and Grudinin \cite{GorodetskyGrudinin} \footnote{This is the discrete sum expression (A15) of Ref. [10], the applicable expression for low frequencies.},
\begin{equation}
S_{df/f} = \frac{8 k_B T^2 \Gamma(R,f,T)}{\pi^2 \gamma(T) R} \left( \frac{1}{n} \frac{dn}{dT} \right)^2
\end{equation}

\begin{equation}
\sigma_{TR}(\tau) = \frac{2 T}{\pi} \sqrt{ \frac{k_B \Gamma(R)}{\gamma(T) R \tau} } \left( \frac{1}{n} \frac{dn}{dT} \right)
\end{equation}

\noindent where $\Gamma$ evaluates to 0.847 for a 1-millimetre-radius CaF$_2$ sphere at low frequency and for the entire temperature range of interest, and $\gamma$ is the thermal conductivity.

\subsection{Brownian boundary noise}

The fluctuation-dissipation theorem says that the power spectral density of fluctuations of coordinate $x$ of a body, at frequency $f$ is
\begin{equation}
S_x(f) = \frac{k_B T}{\pi^2 f^2}|\text{Re}[Y(f)]|
\end{equation}

\noindent where $Y(f) = 2 \pi i f x(f)/F(f)$ is the susceptibility for displacement $x(f)$ under action of the force conjugate to it, $F(f)$. Levin \cite{Levin} noted that
\begin{equation}
|Re[Y(f)]|=\frac{2W_{diss}}{F^2}(f)
\end{equation}

\noindent where $W_{diss}$ is the dissipation when an oscillating force of magnitude $F$ conjugate to the desired displacement and of frequency $f$ is applied. The cycle-averaged dissipation can be written as
\begin{equation}
W_{diss}(f,T) = 2 \pi f U(f) \phi(f,T) \label{MechanicalDissipation}
\end{equation}

\noindent where $U$ is the strain energy in the body at the temporal maximum of the applied force and $\phi(f,T)$ is the loss angle of the material. We assume that $\phi$ is constant with respect to frequency (structural damping). The constant loss-angle model is supported by experimental measurements on ultra-low-expansion material (ULE) Fabry-Perot cavities in the frequency range 1 Hz down to 3 mHz \cite{Numata04}. We thus have
\begin{eqnarray}
S_x(f) = \frac{4}{\pi} \frac{k_B T U(f) \phi(T)}{f F^2} \label{FDTSpectrum}\\
\rightarrow S_{dR/R}(f) = \frac{4}{\pi} \frac{k_B T U \phi(T)}{R^2 f F^2} \label{BrownianSpectrum}
\end{eqnarray}

\noindent where in (\ref{BrownianSpectrum}) we have specialized to the case $x = R$ and divided by the radius squared to get the spectral density of relative radius fluctuations, and the $f$-dependence of $U$ has been dropped (static approximation). Since the relative change of the WGR eigenmode frequency $\nu$ with mode radius $R$ is $d\nu/\nu = dR/R$ \footnote{The effect of the strain-dependence of the refractive index is included in the next subsection; if included here it would cause deviation of this relation from rigorous equality, by $\sim6$ percent.}, the Allan deviation of the relative resonance frequency fluctuations for this $1/f$ spectrum is independent of integration time,
\begin{equation}
\sigma_{BB} = \left[{\frac{8 ln(2)}{\pi}}\right]^{1/2} \frac{\sqrt{k_B T U \phi(T)}}{RF} \label{BrownianAllanDev}
\end{equation}

\noindent We refer to this type of thermal noise as ``Brownian boundary'' (BB) noise.
	
Thus, to calculate the fluctuation of the mode path length, we apply an oscillating force conjugate to a stretching of this path. This is a force along the circumferential optical path in the interior of the resonator, with a transverse spatial distribution identical to the mode intensity profile. A circumferential (ring) load is elastically equivalent to a radial one \footnote{Static equilibrium of the resonator material implies that a tensile ring load $T$ applied at radial position $r$ is necessarily accompanied by a radial compressive load per unit length $X = T/r$}. As the mode is very close to the surface of the resonator, the load can be closely approximated by a surface radial stress distribution with the profile of the 1-dimensional (radial) integral of the mode profile. This is the load that we apply.

The calculation is carried out in two steps. First, the electromagnetic field of a fundamental whispering gallery mode (chosen to be at $\sim 1.9152 \times 10^{14}$ Hz (1565 nm) for concreteness) is simulated using the PDE module of the finite-element package COMSOL \footnote{\itshape COMSOL Multiphysics\upshape, COMSOL, Ab., Tegnergatan 23, SE-111 40 Stockholm, Sweden. Available online: http://www.comsol.com. Version 3.5 used.}, using weak-form partial differential equations following the procedure of Oxborrow \cite{Oxborrow1, Oxborrow2}. The field intensity along the vertical midline of a 2D cross-section of the mode is then fit by a gaussian. The field intensity distribution in the radial direction is also fitted, for use in the elasto-optic calculation described in the next subsection. These fitted profiles are the desired information from the electromagnetic simulation. 

The second step is to calculate the strain energy stored in the resonator when a load with the same profile as the electromagnetic mode intensity is applied to the resonator surface. We do this for a resonator with its axis along the [111] crystal direction, and in the static approximation (i.e. material acceleration is neglected), accurate for low-frequency noise. The calculation is carried out using COMSOL's 3D structural mechanics module. The strain energy is integrated over the body to give $U$ and the applied load is integrated over the surface to give $F$ \footnote{$F$ may also be determined by simple analytic integration as $F = (2 \pi)^{3/2} A R w$, where the applied gaussian load is  $A e^{-(z/(\sqrt{2}w))^2}$.}. Eq.~(\ref{BrownianAllanDev}) then furnishes the Allan deviation. In both the electromagnetic and strain-energy simulations, care is taken to mesh sufficiently densely in the mode region. In the strain energy simulation, deformation patterns observed corresponded well to the applied load profiles, indicating that the force magnitudes used were sufficient. Numerical values of CaF$_2$ properties used in the calculation are listed in Appendix A. 

\subsection{Elasto-optic noise}

The dimensional fluctuation of the resonator also affects the mode frequency indirectly via the the elasto-optic effect, the dependence of the refractive index on strain. Relative fluctuations of different points in the resonator material correspond to fluctuating strains, and the refractive index of the material varies with strain as described by its elasto-optic coefficients \cite{Nye}. We refer to this type of thermal noise as ``elasto-optic'' (EO) noise. This noise term has also recently been considered for multi-layer mirror coatings in Fabry-Perot cavities \cite{Kondratiev}. The effect of the fluctuation of strain component $\epsilon$ on the Allan deviation can be written as 
\begin{equation}
\sigma_{EO} = \frac{1}{\nu} \left( \sigma_{\epsilon} \frac{\partial \nu}{\partial n} \frac{\partial n}{\partial \epsilon} \right) \simeq \frac{\sigma_{\epsilon}n^2p}{2} \label{EOAllanDev1}
\end{equation}

\noindent where $p$ is the elasto-optic coefficient ($p = \frac{\partial (1/\varepsilon)}{\partial \epsilon}$, where $\varepsilon$ is the dielectric constant), and we have used the approximation $\frac{\partial \nu}{\partial n} \simeq -\frac{\nu}{n}$, valid for azimuthal mode index $\gg 1$.

As for the BB noise, we calculate $\sigma_{\epsilon}$ using the direct application of the FDT. To calculate the fluctuation of a given strain component, we apply a load to the resonator conjugate to that strain that is spatially weighted according to the intensity of the mode. We evaluate the strain energy in the resonator resulting from the application of this load, and then obtain the strain fluctuation by using for $x$ in Eq.~(\ref{FDTSpectrum}) the dimension corresponding to the strain in question.

In general there are six strain components $(\epsilon_{11},\epsilon_{22},\epsilon_{33},\epsilon_{12},\epsilon_{13},\epsilon_{23})$ to be considered. We restrict our consideration to volume fluctuations, and we therefore consider only the dilational strain $dV/V = \epsilon_{11}+\epsilon_{22}+\epsilon_{33}$. We expect from the elasto-optic tensor that the effects of shear strain fluctuations will be of similar magnitude.

We perform the dilational calculation by writing the volume of the mode as $\sim 2 \pi^2 r^2 R$ where $r$ is the mode toroid effective minor diameter. Thus the relative fluctuation of the mode volume is 
\begin{equation}
\sigma_{dV/V} =  \sigma_{dR/R} + 2 \sigma_{dr/r} \label{VolumeFluct}
\end{equation}

\noindent $\sigma_{dR/R}$ is given by the calculation of the previous section ($\sigma_{dR/R} = \sigma_{BB}$). We calculate $\sigma_{dr/r}$ by applying a stress distribution $\vec{\Sigma}$ radially directed to the mode center,
\begin{align}
(\Sigma_{\rho},\Sigma_{\theta},\Sigma_z) = &- \Sigma_0 \frac{I(\rho,z)}{I_0} \nonumber \\ 
&\times \left(\frac{\rho-\rho_0}{\sqrt{(\rho - \rho_0)^2+z^2}},0,\frac{z}{\sqrt{(\rho - \rho_0)^2+z^2}}\right)\\
\frac{I(\rho,z)}{I_0} &= Exp \left\{ -\left[ \left( \frac{\rho-\rho_0}{w_{\rho}} \right)^2 + \left(\frac{z}{w_z}\right)^2 \right]\right\}
\end{align}

\noindent where the mode center in cylindrical coordinates $(\rho, \theta, z)$ is the circle $(\rho_0, \theta, 0)$. Then $\sigma_{dr/r}$ is given by
\begin{equation}
\sigma_{dr/r} = \left[ \frac{8 ln(2)}{\pi} \right]^{1/2} \frac{\sqrt{k_B T U \phi(T)}}{r\left[2 \pi \int_{WGR} \left(\Sigma_{\rho} + \Sigma_z \right) \rho d\rho dz \right]}
\end{equation}

\noindent For $r$ in the above relation we use the (averaged) gaussian $1/e$ intensity radius $r = \sqrt{w_{\rho}w_z}$.  The dilational elasto-optic coefficient is given in terms of the cartesian tensor components by $(p_{11}+2p_{12})/3$. Thus from Eq.~(\ref{VolumeFluct}) and Eq.~(\ref{EOAllanDev1}) we have \footnote{The `$+$' sign in Eq.~(\ref{EOAllanDev2}) is understood to indicate the appropriate combination of the terms taking into account any correlation between them that may be present.}
\begin{equation}
\sigma_{EO} \simeq \left(\frac{1}{2} \sigma_{dR/R} + \sigma_{dr/r}\right) \frac{p_{11}+2 p_{12}}{3} n^2 \label{EOAllanDev2}
\end{equation}

\noindent In our calculations we find that $\sigma_{dr/r}/\sigma_{dR/R}$ ranges from 12 to 160 for the resonator sizes considered, and we therefore neglect the first term in Eq.~(\ref{EOAllanDev2}). The contribution of the $\sigma_{dR/R}$ term increases as the WGR size decreases.

\section{Simple estimates}

\noindent \textit{Brownian boundary noise} - We can make a simple estimate of the BB noise of a spherical resonator by considering the strain energy that results from a uniform pressure on its surface. We expect that this will yield a noise smaller than the true value for a spherical WGR, as in this estimate the radius fluctuations will be averaged over the entire spherical surface surface rather than over only the mode region. For a sphere subject to a constant surface pressure $P$ the strain energy $U$ is 
\begin{equation}
U = P \Delta V = P \frac{VP}{\kappa} = \frac{4 \pi}{3} \frac{P^2 R^3}{\kappa}
\end{equation}

\noindent where $\kappa = \left( C_{11} + 2 C_{12} \right)/3$ is the bulk modulus for a cubic crystal with $C_{11}$ and $C_{12}$ elastic constants. The applied force is $F = 4 \pi R^2 P$, thus we estimate
\begin{eqnarray}
\tilde{S}_{dR/R}(f) = \frac{1}{3 \pi^2} \frac{k_B T \phi(T)}{\kappa R^3 f}\\
\tilde{\sigma}_{BB}(\tau) = \left[ \frac{(2/3)ln(2)}{\pi} \right]^{1/2} \sqrt{ \frac{k_B T \phi(T)}{\kappa R^3} } \label{PhiUniformSphere}
\end{eqnarray}

\noindent For a 1-millimetre-radius sphere at 5.5 K, with $\phi(5 K) = 2 \times 10^{-8}$ \cite{Nawrodt_CaF2_Q}, this gives $\sigma_{BB} = 3 \times 10^{-17}$ for CaF$_2$ ($\kappa$ = 90 GPa), a factor of 2.5 smaller than the value determined by more detailed calculations below.\\

\noindent \textit{Elasto-optic noise} - We can model the WGM volume as a tube of length $2 \pi R$ and radius $r$. We estimate $r$ from $V_m = 2 \pi^2 R r^2$, where the mode volume $V_m$ is given by \cite{Braginsky89} $V_m \simeq 3.4 \pi^{3/2} (\lambda/n)^{7/6} R^{11/6}$, where $\lambda$ is the optical wavelength,
\begin{equation}
r \simeq 0.335 (\lambda/n)^{7/12} R^{5/12}
\end{equation}

\noindent We calculate the strain energy that results from a uniform pressure $P$ applied to the cylindrical surface of the tube, approximating the WGR material as isotropic. Applying the plane-strain condition, the non-zero stresses in the tube are $\sigma_{rr} = \sigma_{\theta \theta} = -P$, $\sigma_{zz} = - 2 \mu P$, where $\mu$ is Poisson's ratio, and the resulting strain energy is
\begin{equation}
U = \pi^2 r^2 R P^2 \left[\frac{3}{3\kappa + G}\right]
\end{equation} 

\noindent where $G$ is the shear modulus. The applied force is $F = (2 \pi)^2 r R P$, and we get
\begin{eqnarray}
&\tilde{\sigma}_{dr/r}(\tau) = \frac{\sqrt{6 ln(2)}}{2 \pi^3/2}\left[\frac{k_B T \phi}{r^2 R (3\kappa + G)}\right]^{1/2} \nonumber \\ 
&= 0.55 \left(\frac{n}{\lambda}\right)^{7/12}R^{-11/12}\left(\frac{k_B T \phi}{3\kappa + G} \right)^{1/2}\\
\tilde{\sigma}_{EO}(\tau) &\simeq 0.55  \left(\frac{n^{31/12}}{\lambda^{7/12}R^{11/12}}\right) \left(\frac{k_B T \phi}{3\kappa + G} \right)^{1/2} \frac{p_{11} + 2 p_{12}}{3}
\end{eqnarray}

\noindent For a 1-mm-radius resonator, at $\lambda$ = 1.56 $\mu$m and with $G$ = 45 GPa, this numerically evaluates at 5.5 K to $7 \times 10^{-16}$, close to the more accurately calculated result below.

\section{Results}

We perform the calculation for sphere and disk radii ranging from 0.1 mm to 10 mm, and for disk surface radii of curvature ranging from 0.1 mm to 10 mm. The disk shapes used are shown in Fig.~\ref{fig:Simulation}. 
To numerically evaluate the Allan deviation with equations (\ref{BrownianAllanDev}) and (\ref{EOAllanDev2}) we must know the material loss angle $\phi$. Nawrodt et. al. \cite{Nawrodt_CaF2_Q} measured the $Q$ of mechanical modes of a pure crystalline CaF$_2$ cylinder as a function of temperature. In the cylinder's fundamental drum mode, material motion is concentrated near the cylinder axis and away from the experimental support. We therefore take the reported $Q$ of the cylinder's fundamental drum mode as representative of the bulk material loss and therefore as the limiting value achievable for losses. The $Q$ of this mode ranged from $10^{7}$ to $5 \times 10^8$ between 5 K and 300 K. Using this loss angle $\phi = 1/Q = (2 \times 10^{-8}, 5 \times 10^{-8})$ (5 K, 300 K values), we obtain the Allan deviation values shown in Table~\ref{table_4}. Uncertainty in $\phi$ is the principal uncertainty in our calculation. For different values of $\phi$ the BB and EO noise terms scale as $\phi^{1/2}$.

\begin{table}[b!]
\caption{\label{table_4} Calculated Brownian boundary and elasto-optic noise for CaF$_2$ spheres and disks at 300 K and 5 K. $R$: radius; $S$: surface radius of curvature.}
\begin{center}
\begin{ruledtabular}
\begin{tabular}{ m{1.2cm} m{1.4cm} m{1.4cm} m{1.4cm} m{1.4cm} }
 $R$, $S$ & $\sigma_{BB}$ & $\sigma_{EO}$ & $\sigma_{BB}$ & $\sigma_{EO}$ \\
  (mm) & RT & RT & 5 K & 5 K \\ \hline 
 {\smallskip}\\
 \multicolumn{5}{l}{\underline{Spheres}} \\ 
 0.1 & 2$\times$10$^{-14}$ & 9$\times$10$^{-14}$ & 2$\times$10$^{-15}$ & 8$\times$10$^{-15}$ \\
 1.0 & 8$\times$10$^{-16}$ & 1$\times$10$^{-14}$ & 7$\times$10$^{-17}$ & 1$\times$10$^{-15}$ \\
 10 & 3$\times$10$^{-17}$ & 2$\times$10$^{-15}$ & 3$\times$10$^{-18}$ & 1$\times$10$^{-16}$\\ 
 {\smallskip}\\
 \multicolumn{5}{l}{\underline{Disks}}   \\ 
 0.1, 0.15 & 2$\times$10$^{-14}$ & 9$\times$10$^{-14}$ & 2$\times$10$^{-15}$ & 8$\times$10$^{-15}$ \\
 1.0, 0.15 & 9$\times$10$^{-16}$ & 1$\times$10$^{-14}$ & 8$\times$10$^{-17}$ & 1$\times$10$^{-15}$ \\
 10, 0.15 & 5$\times$10$^{-17}$ & 2$\times$10$^{-15}$ & 4$\times$10$^{-18}$ & 2$\times$10$^{-16}$ \\ 
  1.0, 0.1 & 9$\times$10$^{-16}$ & 1$\times$10$^{-14}$ & 8$\times$10$^{-17}$ & 1$\times$10$^{-15}$\\
 1.0, 1.0 & 9$\times$10$^{-16}$ & 1$\times$10$^{-14}$ & 8$\times$10$^{-17}$ & 1$\times$10$^{-15}$\\
 1.0, 10 & 9$\times$10$^{-16}$ & 1$\times$10$^{-14}$ & 7$\times$10$^{-17}$ & 1$\times$10$^{-15}$\\ 
\end{tabular}
\end{ruledtabular}
\end{center}
\end{table}

\begin{figure}[t] 
\begin{center}
\includegraphics[width=3.3in]{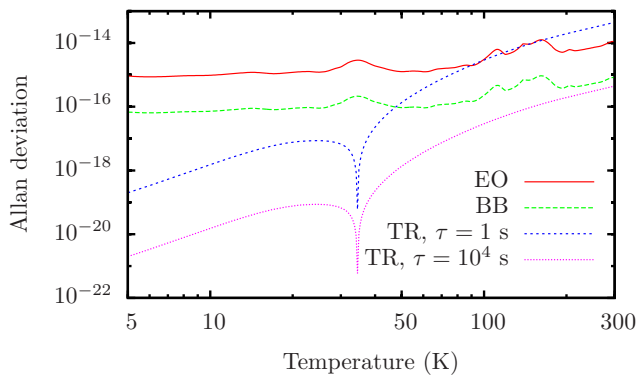}
\caption{\label{fig:AllanTemperature} (Color online) Thermal noise for a 1-millimetre-radius CaF$_2$ sphere from 5.5 K to 300 K. TR: thermorefractive noise; BB: Brownian boundary noise; EO: elasto-optic noise. There is a zero in the first-order thermorefractive noise at $\sim$33 K due to the zero-crossing of $dn/dT$.}
\end{center}
\end{figure}

Figure~\ref{fig:AllanTemperature} shows the different thermal noise terms for a millimetre-sized resonator. At room temperature and short time scales, the dominant thermal noise contribution is the thermorefractive noise, in agreement with experiments \cite{GorodetskyGrudinin, MPQ_MgF2_TRNoise}. The BB and EO noises vary with temperature as $[T\phi(T)]^{1/2}$, while the temperature-dependence of the thermorefractive noise is the much stronger function $T(1/n)(dn/dT) \sqrt{\gamma(T)}$. Thus the BB and EO noise terms become relatively more important with decreasing temperature. The EO noise is an order of magnitude larger than the BB noise for millimetre-sized WGRs, and becomes the dominant noise component at low temperatures and long time scales. 

\begin{figure}[t] 
\begin{center}
\includegraphics[width=3.3in]{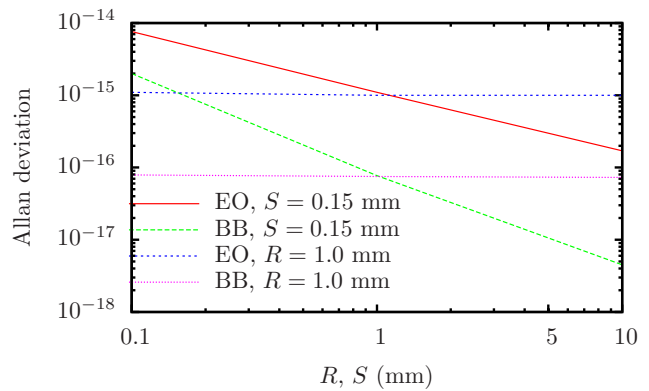}
\caption{\label{fig:AllanSize}(Color online) 5.5 K Allan deviation due to Brownian boundary noise and elasto-optic noise of a CaF$_2$ disk WGM, as a function of disk radius $R$ and surface vertical radius of curvature $S$. BB, $R$ = 1.0 mm: Brownian boundary noise as a function of $S$, for $R$ = 1.0 mm. $\phi$ = $2 \times 10^{-8}$.}
\end{center}
\end{figure}

Figure~\ref{fig:AllanSize} shows the scaling of the calculated noise terms with disk radius and surface radius of curvature. We see that within the range of values considered, the noise is essentially independent of surface radius of curvature, and the ratio of EO to BB noise scales approximately as $R^{0.5}$  ($\sigma_{EO} \propto \,\, \sim \! R^{-0.8}$, $\sigma_{BB} \propto \,\, \sim \! R^{-1.3}$). For a sphere the scaling is similar ($\sigma_{EO} \propto \,\, \sim R^{-0.9}$ , $\sigma_{BB} \propto \,\, \sim \! R^{-1.4}$). This scaling agrees with the simple estimates in the previous section, for which $\sigma_{EO} \propto \, R^{-0.9}$, $\sigma_{BB} \propto \, R^{-1.5}$. The thermorefractive noise depends on mode volume in the same way as does the EO noise, and should thus have a similar dependence on $R$.

Thus we see that lower noise levels are achieved by going to lower temperatures and larger resonators, with extrapolated noise levels at the $10^{-17}$ level for a 10-mm-radius sphere at 25 mK and for a 10-cm-radius sphere at 1.7 K, where we have used the 5 K value of $\phi$.

\section{Conclusion}

We have calculated the thermal noise limit to the frequency stability of crystalline CaF$_2$ whispering-gallery resonators set by dimensional fluctuations, by direct application of the fluctuation-dissipation theorem and using the widely-applied structural damping model. We have identified a new source of thermal noise in whispering gallery resonators, due to the elasto-optic effect. This noise is smaller than the thermorefractive noise at room temperature, so it has not been observed in previous room-temperature experiments. At temperatures sufficiently below room temperature (e.g. 50 K), the elasto-optic noise dominates. The total thermal noise is reduced by increasing the resonator size and reducing the temperature. WGRs have already been operated at cryogenic temperatures \cite{ArcizetCryo, SchliesserEtAl}, and therefore an experimental study of the noise at low temperatures should be feasible. One important result of such studies will be a determination of the loss angle $\phi$, which can at present only be estimated. For achieving extremely low total noise, it would be favourable to operate large resonators ($R \ge$ 50 mm) at temperatures $\le$ 100 mK, where a noise level $\le$ $4 \times 10^{-17}$ is estimated from the present work.

\begin{acknowledgments}

This work was performed in the framework of project AO/1-5902/09/D/JR of the European Space Agency. We thank J. de Vicente and I. Zayer for support.

\end{acknowledgments}

\appendix

\section{Material properties}

\begin{table}[!h]
\caption{\label{table_A1} CaF$_2$ material properties. Property values at intermediate temperatures may be found in the cited references.}
\renewcommand\tabcolsep{5pt}
\begin{center}
\begin{tabular}{ l c  c  c }
\toprule
 \multicolumn{2}{ c }{Property}                           & Value                                & Ref.                                      \\ \colrule\noalign{\smallskip}
 Refractive index ($n$)                 &                  & 1.43                                  & \cite{Leviton_CaF2_TRIndex}\\ \noalign{\smallskip}
 Elastic constants:                           & $C_{11}$ & 164 GPa                            & \cite{HuffmanNorwood}          \\
                                                        & $C_{12}$ & 53.0 GPa                          &                                                \\
                                                        & $C_{44}$ & 33.7 GPa                          &                                             \\ \noalign{\smallskip}
Thermal conductivity ($\gamma$): & 300 K      & 9.71 Wm$^{-1}$K$^{-1}$  & \cite{Slack_Conductivity}        \\
                                                        & 5.5 K        & 1200 Wm$^{-1}$K$^{-1}$ &                                             \\ \noalign{\smallskip}
Thermorefractive index                  & 295 K       & $8.4 \times 10^{-6}$         & \cite{Leviton_CaF2_TRIndex}  \\
 ($|(1/n)(dn/dT)|$):                          & 5.5 K        & $1.5 \times 10^{-9}$         &                                             \\ \noalign{\smallskip}
 Loss angle ($\phi$):                       & 300 K       & $5 \times 10^{-8}$            & \cite{Nawrodt_CaF2_Q}            \\
                                                       & 5.5 K        & $2 \times 10^{-8}$             &                                             \\ \noalign{\smallskip}
 Elasto-optic constants:                  & $p_{11}$  &  0.039                                & \cite{Burnett_POCoefficients}   \\
                                                       & $p_{12}$  &  0.223                                &                                                  \\
                                                       & $p_{44}$  &  0.051                                &                                        \\ \noalign{\smallskip}
Thermal expansion                        & 295 K &  $1.89 \times 10^{-5}$ K$^{-1}$ & \cite{BatchelderSimmons}\\
 coefficient ($\alpha$):                     & 5.5 K &  $6.4 \times 10^{-10}$ K$^{-1}$ & \footnote{Extrapolation from 20 K data using $T^3$ dependence.}\\ \botrule
\end{tabular}
\end{center}
\end{table}

\section{Calculation results}

We tabulate here more detailed results of the numerical calculation.

\subsubsection{Electromagnetic modeling}

\clearpage

\begin{table}[!h]
\caption{\label{table_B1} Modeled fundamental electromagnetic modes: sphere}
\begin{ruledtabular}
\begin{tabular}{D{.}{.}{3,1} D{.}{.}{4,1} D{.}{.}{2,4} D{.}{.}{3,5}}
\multicolumn{1}{c}{Sphere} & \multicolumn{1}{c}{Mode}   & \multicolumn{1}{c}{Mode} & \multicolumn{1}{c}{Mode polar}   \\ 
\multicolumn{1}{c}{radius} & \multicolumn{1}{c}{azimuthal} & \multicolumn{1}{c}{frequency} & \multicolumn{1}{c}{intensity 1/e$^2$} \\
\multicolumn{1}{c}{(mm)} &  \multicolumn{1}{c}{index} & \multicolumn{1}{c}({$10^{14}$ Hz}) & \multicolumn{1}{c}{half-width ($\mu$m)} \\ \hline
 0.1 & 559 & 1.9164 & 4.25\\
 1 & 5706 & 1.9152 & 13.5 \\
 10 & & 1.9149 & 42.0 \\ 
\end{tabular}
\end{ruledtabular}
\end{table}

\begin{table}[!h]
\caption{\label{table_B2} Modeled fundamental electromagnetic modes: disk}
\begin{center}
\begin{ruledtabular}
\begin{tabular}{D{.}{.}{2,1} D{.}{.}{2,2} D{.}{.}{5,0} D{.}{.}{2,4} D{.}{.}{2,4}}
 \multicolumn{1}{c}{Disk} & \multicolumn{1}{c}{Vertical radius} & \multicolumn{1}{c}{Mode} & \multicolumn{1}{c}{Mode} & \multicolumn{1}{c}{Mode polar} \\
  \multicolumn{1}{c}{radius} & \multicolumn{1}{c}{of curvature} & \multicolumn{1}{c}{azimuthal} & \multicolumn{1}{c}{frequency} & \multicolumn{1}{c}{intensity 1/e$^2$} \\
  \multicolumn{1}{c}{(mm)} & \multicolumn{1}{c}{(mm)} & \multicolumn{1}{c}{index} & \multicolumn{1}{c}({10$^{14}$ Hz}) & \multicolumn{1}{c}{half-width ($\mu$m)} \\ \hline
 0.1 & 0.15 & 559 & 1.9156 & 4.6 \\
 1 & 0.15 & 5706 & 1.9152 & 8.2 \\
 10 & 0.15 & 57320 & 1.9151 & 14.1 \\
 1 & 0.10 & 5707 & 1.9151 & 6.7 \\
 1 & 1 & 5707 & 1.9151 & 11.3 \\
 1 & 10 & 5708 & 1.9155  & 13.0 \\ 
\end{tabular}
\end{ruledtabular}
\end{center}
\end{table}

\subsubsection{Strain energy calculation}

\begin{table}[!h]
\caption{\label{table_1} BB noise calculation for CaF$_2$ spheres. $R$: sphere radius, $w_z$: mode 1/e$^2$ intensity half-width, $U$: strain energy $10^6$ N/m$^2$ peak load, $F$: total force, $\sigma_{dR/R}$: Allan deviation for $\phi = 2 \times 10^{-8}$.}
\begin{center}
\begin{ruledtabular}
\begin{tabular}{ c D{.}{.}{2,2} l D{.}{.}{2,4} c }
 \multicolumn{1}{c}{$R$} & \multicolumn{1}{c}{$w_z$} & \multicolumn{1}{c}{$U$} & \multicolumn{1}{c}{$F$}  & \multicolumn{1}{c}{$\sigma_{dR/R}$}\\
 \multicolumn{1}{c}{(mm)} &  \multicolumn{1}{c}{($\mu$m)} & \multicolumn{1}{c}{(J)} & \multicolumn{1}{c}{(N)} & at 5.5 K \\ \hline
 0.1 & 4.25 & 3.2$\times$10$^{-13}$ & 0.0047 & 2.0$\times$10$^{-15}$ \\
 1 & 13.5 & 4.4$\times$10$^{-11}$ & 0.15 & 7.2$\times$10$^{-17}$ \\
 10 & 42.0 & 5.3$\times$10$^{-9}$ & 4.7 & 2.6$\times$10$^{-18}$ \\
\end{tabular}
\end{ruledtabular}
\end{center}
\end{table}

\begin{table}[!h]
\caption{\label{table_3} BB noise calculation for CaF$_2$ disks. $R$: disk radius, $S$: surface radius of curvature.}
\begin{center}
\begin{ruledtabular}
\begin{tabular}{l D{.}{.}{2,2} l D{.}{.}{2,4} c }
\multicolumn{1}{c}{$R$, $S$} & \multicolumn{1}{c}{\text{$w_z$}} & \multicolumn{1}{c}{$U$} &\multicolumn{1}{c}{$F$} & \multicolumn{1}{c}{\text{$\sigma_{dR/R}$}} \\
\multicolumn{1}{c}{(mm)} & \multicolumn{1}{c}{\text{($\mu$m)}} & \multicolumn{1}{c}{\text{(J)}} & \multicolumn{1}{c}{\text{(N)}} & at 5.5 K \\ \hline
 0.1, 0.15 & 4.6 & 3.8$\times$10$^{-13}$ & 0.0051 & 2.0$\times$10$^{-15}$ \\
 1, 0.15 & 8.2 & 1.9$\times$10$^{-11}$ & 0.091 & 7.7$\times$10$^{-17}$ \\
 10, 0.15 & 14.1 & 1.8$\times$10$^{-9}$ & 1.6 & 4.5$\times$10$^{-18}$ \\
 1, 0.1 & 6.7 & 1.3$\times$10$^{-11}$ & 0.075 & 7.9$\times$10$^{-17}$ \\
 1, 1 & 11.3 & 3.3$\times$10$^{-11}$ & 0.13 & 7.5$\times$10$^{-17}$ \\
 1, 10 & 13.0 & 4.2$\times$10$^{-11}$ & 0.14 & 7.3$\times$10$^{-17}$ \\
\end{tabular}
\end{ruledtabular}
\end{center}
\end{table}

\begin{table}[!h]
\caption{\label{} EO noise calculation for CaF$_2$ spheres. $w_\rho$: mode radial 1/e$^2$ intensity half-width, $\rho_0$: radial position of mode center.}
\begin{center}
\begin{ruledtabular}
\begin{tabular}{ c c D{.}{.}{1,4} m{1.4cm} D{.}{.}{2,4} m{1.35cm} }
$R$ & $w_z$, $w_{\rho}$ & \multicolumn{1}{c}{$\rho_0$} & \multicolumn{1}{c}{$U$} & \multicolumn{1}{c}{$F$} & \multicolumn{1}{c}{$\sigma_{dr/r}$}\\
\multicolumn{1}{c}{(mm)} &  ($\mu$m) & \multicolumn{1}{c}{(mm)} & \multicolumn{1}{c}{(J)} & \multicolumn{1}{c}{(N)} & at 5.5 K \\ \hline
 0.1 & 4.25, 1.1 & 0.0986 & 7.1$\times$10$^{-14}$ & 0.0088 & 2.3$\times$10$^{-14}$ \\
 1 & 13.5, 2.5 & 0.996 & 4.6$\times$10$^{-11}$ & 0.65 & 2.9$\times$10$^{-15}$ \\
 10 & 42.0, 5.0 & 9.99 & 2.3$\times$10$^{-8}$ & 41 & 4.2$\times$10$^{-16}$ \\
\end{tabular}
\end{ruledtabular}
\end{center}
\end{table}

\begin{table}[!h]
\caption{\label{t} EO noise calculation for CaF$_2$ disks.}
\begin{center}
\begin{ruledtabular}
\begin{tabular}{l  c D{.}{.}{1,3} c D{.}{.}{2,4} c }
\multicolumn{1}{c}{$R$, $S$} & \multicolumn{1}{c}{$w_z$, $w_{\rho}$} & \multicolumn{1}{c}{$\rho_0$}  & \multicolumn{1}{c}{$U$} & \multicolumn{1}{c}{$F$} & \multicolumn{1}{c}{\text{$\sigma_{dr/r}$}} \\
\multicolumn{1}{c}{(mm)}& \multicolumn{1}{c}{\text{($\mu$m)}} & \multicolumn{1}{c}{(mm)} & \multicolumn{1}{c}{\text{(J)}} & \multicolumn{1}{c}{\text{(N)}} & at 5.5 K \\ \hline
 0.1, 0.15 & 4.6, 1.1  & 0.986 & 8.8$\times$10$^{-14}$ & 0.0094 & 2.3$\times$10$^{-14}$ \\
 1, 0.15 & 8.2, 2.5 & 0.997 & 1.2$\times$10$^{-11}$ & 0.39 & 3.3$\times$10$^{-15}$ \\
 10, 0.15 & 14.1, 5.3  & 0.999 & 1.5$\times$10$^{-9}$ & 14 & 5.2$\times$10$^{-16}$ \\
  1, 0.1 & 6.7, 2.5  & 0.997 & 6.7$\times$10$^{-12}$ & 0.31& 3.3$\times$10$^{-15}$ \\
 1, 1 & 11.3, 2.5  & 0.997 & 2.9$\times$10$^{-11}$ & 0.53 & 3.1$\times$10$^{-15}$ \\
 1, 10 & 13.0, 2.5  & 0.997 & 4.2$\times$10$^{-11}$ & 0.61 & 3.0$\times$10$^{-15}$ \\
\end{tabular}
\end{ruledtabular}
\end{center}
\end{table}

\clearpage

\bibliography{ThermalNoiseRefs_Rev}

\end{document}